\documentclass[aps,prl,preprint,superscriptaddress,floatfix]{revtex4-1}
\usepackage[bookmarksnumbered,unicode]{hyperref}
\usepackage{graphicx}
\usepackage{amsmath,amssymb}
\usepackage{bm}
\usepackage{multirow}
\usepackage{color}

\def \d {\mathrm{d}}

\begin{document}

%\preprint{}

\title{Surface localization of the dineutron in \texorpdfstring{$^{11}$}{11}Li}


%\author{}
%\thanks{}
%\affiliation{}


\def \cns      {Center for Nuclear Study, University of Tokyo, Hongo 7-3-1, Bunkyo, Tokyo 113-0033, Japan}
\def \rnc      {RIKEN Nishina Center, Hirosawa 2-1, Wako, Saitama 351-0198, Japan}
\def \saclay   {D\'{e}partement de Physique Nucl\'{e}aire, IRFU, CEA, Universit\'{e} Paris-Saclay, F-91191 Gif-sur-Yvette, France}
\def \tud      {Institut f\"{u}r Kernphysik, Technische Universit\"{a}t Darmstadt, D-64289 Darmstadt, Germany}
\def \peking   {School of Physics and State Key Laboratory of Nuclear Physics and Technology, Peking University, Beijing 100871, China}
\def \orsay    {IPN Orsay, Universit\'{e} Paris Sud, IN2P3-CNRS, F-91406 Orsay Cedex, France}
\def \caen     {LPC Caen, ENSICAEN, Universit\'{e} de Caen Normandie, CNRS/IN2P3, F-14050 Caen Cedex, France}
\def \tohoku   {Department of Physics, Tohoku University, Aramaki Aza-Aoba 6-3, Aoba, Sendai, Miyagi 980-8578, Japan}
\def \miyazaki {Department of Applied Physics, University of Miyazaki, Gakuen-Kibanadai-Nishi 1-1, Miyazaki 889-2192, Japan}
\def \ehwa     {Department of Physics, Ehwa Womans University, Seoul 03760, Republic of Korea}
\def \ut       {Department of Physics, University of Tokyo, Hongo 7-3-1, Bunkyo, Tokyo 113-0033, Japan}
\def \titech   {Department of Physics, Tokyo Institute of Technology, 2-12-1 O-Okayama, Meguro, Tokyo 152-8551, Japan}
\def \atomki   {Institute for Nuclear Research, Hungarian Academy of Sciences (MTA Atomki), P.O. Box 51, H-4001 Debrecen, Hungary}
\def \kyoto    {Department of Physics, Kyoto University, Kitashirakawa, Sakyo, Kyoto 606-8502, Japan}
\def \kyushu   {Department of Physics, Kyushu University, Nishi, Fukuoka 819-0395, Japan}
\def \tum      {Physik Department, Technische Universit\"{a}t M\"{u}nchen, D-85748 Garching, Germany}
\def \rcnp     {Research Center for Nuclear Physics, Osaka University, 10-1 Mihogaoka, Ibaraki, Osaka 567-0047, Japan}
\def \ikp      {Institut f\"{u}r Kernphysik, Technische Universit\"{a}t Darmstadt, D-64289 Darmstadt, Germany}
\def \cenm     {Center for Extreme Nuclear Matters, Korea University, Seoul 02841, Republic of Korea}
\def \cens     {Center for Exotic Nuclear Studies, Institute for Basic Science (IBS), Daejeon 34126, Korea}
\def \ocu      {Department of Physics, Osaka City University, Osaka 558-8585, Japan}
\def \nit      {Tokuyama College, National Institute of Technology, Yamaguchi 745-8585, Japan}
\def \york     {Department of Physics, University of York, Heslington, York YO10 5DD, United Kingdom}
\def \lbnl     {Nuclear Science Division, Lawrence Berkeley National Laboratory, Berkeley, California 94720, USA}
\def \cpr      {Cluster for Pioneering Research, RIKEN, Hirosawa 2-1, Wako, Saitama 351-0198, Japan}

\author{Y.~Kubota}\email{kubota@ribf.riken.jp}\altaffiliation[Present address: ]{\ikp}\affiliation{\rnc}\affiliation{\cns}
\author{A.~Corsi}\affiliation{\saclay}
\author{G.~Authelet}\affiliation{\saclay}
\author{H.~Baba}\affiliation{\rnc}
\author{C.~Caesar}\affiliation{\tud}
\author{D.~Calvet}\affiliation{\saclay}
\author{A.~Delbart}\affiliation{\saclay}
\author{M.~Dozono}\affiliation{\cns}
\author{J.~Feng}\affiliation{\peking}
\author{F.~Flavigny}\altaffiliation[Present address: ]{\caen}\affiliation{\orsay}
\author{J.-M.~Gheller}\affiliation{\saclay}
\author{J.~Gibelin}\affiliation{\caen}
\author{A.~Giganon}\affiliation{\saclay}
\author{A.~Gillibert}\affiliation{\saclay}
\author{K.~Hasegawa}\affiliation{\tohoku}
\author{T.~Isobe}\affiliation{\rnc}
\author{Y.~Kanaya}\affiliation{\miyazaki}
\author{S.~Kawakami}\affiliation{\miyazaki}
%\author{D.~Kim}\altaffiliation[Present address: ]{\cenm}\affiliation{\ehwa} % old
\author{D.~Kim}\altaffiliation[Present address: ]{\cens}\affiliation{\ehwa} % updated
% \author{Y.~Kikuchi}\altaffiliation[Present address: ]{\nit}\affiliation{\rnc}\affiliation{\ocu}
\author{Y.~Kikuchi}\affiliation{\nit}\affiliation{\rnc}\affiliation{\ocu}
% \author{Y.~Kikuchi}\affiliation{\rnc}\affiliation{\ocu}
\author{Y.~Kiyokawa}\affiliation{\cns}
\author{M.~Kobayashi}\affiliation{\cns}
% \author{N.~Kobayashi}\altaffiliation[Present address: ]{\rcnp}\affiliation{\ut}
\author{N.~Kobayashi}\affiliation{\ut}
\author{T.~Kobayashi}\affiliation{\tohoku}
\author{Y.~Kondo}\affiliation{\titech}
% \author{Z.~Korkulu}\altaffiliation[Present address: ]{\cens}\affiliation{\atomki}
\author{Z.~Korkulu}\affiliation{\cens}\affiliation{\atomki}
\author{S.~Koyama}\affiliation{\ut}
\author{V.~Lapoux}\affiliation{\saclay}
\author{Y.~Maeda}\affiliation{\miyazaki}
\author{F.~M.~Marqu\'{e}s}\affiliation{\caen}
\author{T.~Motobayashi}\affiliation{\rnc}
\author{T.~Miyazaki}\affiliation{\ut}
\author{T.~Nakamura}\affiliation{\titech}
\author{N.~Nakatsuka}\affiliation{\kyoto}
\author{Y.~Nishio}\affiliation{\kyushu}
\author{A.~Obertelli}\altaffiliation[Present address: ]{\ikp}\affiliation{\saclay}
\author{K.~Ogata}\affiliation{\rcnp}\affiliation{\ocu}
\author{A.~Ohkura}\affiliation{\kyushu}
\author{N.~A.~Orr}\affiliation{\caen}
\author{S.~Ota}\affiliation{\cns}
\author{H.~Otsu}\affiliation{\rnc}
\author{T.~Ozaki}\affiliation{\titech}
% \author{V.~Panin}\altaffiliation[Present address: ]{\saclay}\affiliation{\rnc}
\author{V.~Panin}\affiliation{\rnc}
\author{S.~Paschalis}\altaffiliation[Present address: ]{\york}\affiliation{\tud}
% \author{S.~Paschalis}\affiliation{\tud}
\author{E.~C.~Pollacco}\affiliation{\saclay}
\author{S.~Reichert}\affiliation{\tum}
\author{J.-Y.~Rouss\'{e}}\affiliation{\saclay}
\author{A.~T.~Saito}\affiliation{\titech}
\author{S.~Sakaguchi}\affiliation{\kyushu}
\author{M.~Sako}\affiliation{\rnc}
% \author{C.~Santamaria}\altaffiliation[Present address: ]{\lbnl}\affiliation{\saclay}
\author{C.~Santamaria}\affiliation{\saclay}
\author{M.~Sasano}\affiliation{\rnc}
\author{H.~Sato}\affiliation{\rnc}
\author{M.~Shikata}\affiliation{\titech}
\author{Y.~Shimizu}\affiliation{\rnc}
\author{Y.~Shindo}\affiliation{\kyushu}
% \author{L.~Stuhl}\altaffiliation[Present address: ]{\atomki}\affiliation{\rnc}
% \author{L.~Stuhl}\altaffiliation[Present address: ]{\cens}\affiliation{\rnc}
\author{L.~Stuhl}\affiliation{\cens}\affiliation{\rnc}
\author{T.~Sumikama}\affiliation{\tohoku}
\author{Y.~L.~Sun}\altaffiliation[Present address: ]{\ikp}\affiliation{\saclay}
\author{M.~Tabata}\affiliation{\kyushu}
\author{Y.~Togano}\affiliation{\titech}
\author{J.~Tsubota}\affiliation{\titech}
\author{Z.~H.~Yang}\affiliation{\rnc}
\author{J.~Yasuda}\affiliation{\kyushu}
\author{K.~Yoneda}\affiliation{\rnc}
% \author{J.~Zenihiro}\altaffiliation[Present address: ]{\kyoto}\affiliation{\rnc}
\author{J.~Zenihiro}\affiliation{\rnc}
\author{T.~Uesaka}\affiliation{\rnc}\affiliation{\cpr}

\date{\today}

\begin{abstract}
  The formation of a dineutron in the nucleus $^{11}$Li is found to be localized to the surface region.
  The experiment measured the intrinsic momentum of the struck neutron in $^{11}$Li via the $(p,pn)$ knockout reaction at 246~MeV/nucleon.
  The correlation angle between the two neutrons is, for the first time, measured as a function of the intrinsic neutron momentum.
  A comparison with reaction calculations reveals the localization of the dineutron at $r\sim3.6$~fm.
  The results also support the density dependence of dineutron formation as deduced from Hartree-Fock-Bogoliubov calculations for nuclear matter.
\end{abstract}

\pacs{}
%\keywords{}

\maketitle

The structures of fermionic many-body systems are often characterized by two-particle correlations~\cite{PhysRev.114.1377,PhysRevLett.72.40}.
In nuclei, Bardeen-Cooper-Schrieffer-(BCS)-like pairings between two nucleons are known to stabilize nuclei similar to the behavior in superconducting materials~\cite{PhysRev.110.936}.
On the other hand, a completely different type of two-neutron correlation, \textit{dineutron} correlation, has attracted much attention in nuclear physics and other relevant fields~\cite{BERTSCH1991327}.
In the 1970's, A.~B.~Migdal postulated the dineutron as a spatially compact two-neutron pair~\cite{Migdal1973}.
Behind the formation of a dineutron, quantum interference between odd and even parity orbits plays a crucial role~\cite{PhysRevC.29.1091}.
The appearance of the dineutron correlation is presumed to be a key ingredient to elucidate the stabilities and exotic structures of nuclei near and beyond the neutron drip-line nuclei~\cite{Hansen_1987}.

The dineutron correlation in a finite nuclear system may provide insight into dilute neutron-rich matter found in the inner crust of neutron stars. 
Recently, Hartree-Fock-Bogoliubov calculations predicted that the correlation length of two neutron pairs drastically changes with the matter density.
Consequently, the dineutron correlation should appear below the saturation density $\rho_0$ of $10^{-4}\lesssim\rho/\rho_0\lesssim0.5$~\cite{PhysRevC.73.044309}.
Hence, laboratories studies on finite nuclei should elucidate information about the existence and properties of dineutrons.

Studies on the formation and the density dependence of $^{11}$Li as well as $^6$He~\cite{PhysRevLett.82.4996,PhysRevC.60.044605,PhysRevLett.87.042501} are crucial because $^{11}$Li has a peculiar structure, which is known as the halo.
That is, the matter density gradually varies from the saturated core to the very low-density tail where only valence neutrons exist.
If the spatial distribution of the dineutron can be experimentally determined, the density-dependent properties of dineutron may be revealed.

Many studies have examined the dineutron correlation in $^{11}$Li using the transfer~\cite{,TANIHATA1992307} or the Coulomb breakup reactions~\cite{PhysRevC.48.326,PhysRevC.54.1589,Marques2000219}.
One pioneering study measured the electric dipole ($E1$) response of $^{11}$Li using the Coulomb breakup reaction~\cite{PhysRevLett.70.730,PhysRevC.59.1252,PhysRevLett.96.252502}.
The $E1$ cluster sum rule value is directly related to the opening angle $\langle\theta_{12}\rangle$ of the two valence neutrons with respect to the core~\cite{BERTSCH1991327}.
In the absence of a correlation, the $\langle\theta_{12}\rangle$ value is 90$^\circ$ and it becomes smaller as the dineutron correlation becomes stronger.
Hence, $\langle\theta_{12}\rangle$, which can be determined by the $E1$ cluster sum rule value, provides a good measure of the dineutron correlation strength.
A latest result from RIKEN~\cite{PhysRevLett.96.252502} showed that the $E1$ strength for $^{11}$Li integrated over the relative energy region of $E_\mathrm{rel} \leq 3$~MeV is consistent with $\langle\theta_{12}\rangle = 48^{+14}_{-18}$~degrees, indicating that $^{11}$Li has a strong dineutron correlation.
Recent theoretical works have reevaluated the opening angle from the same data, and extracted larger values by including the excitation of the $^9$Li core~\cite{PhysRevC.87.034606}, the distance between the $^9$Li core and a halo neutron~\cite{PhysRevC.76.051602}, and the effect of the final state interaction (FSI) on the $B(E1)$ distribution~\cite{PhysRevC.81.044308}.

In addition, the dineutron correlation has been investigated by measuring the charge radius~\cite{PhysRevLett.93.113002,PhysRevLett.96.033002}.
$^{11}$Li has a larger charge radius than the $^9$Li, indicating the two halo neutrons in $^{11}$Li are distributed on one side forming the compact dineutron and not distributed symmetrically.

The above two methods average dineutron information over the whole volume of the nucleus.
Therefore, neither the location nor the density dependence of the dineutron formation can be determined.
However, the spatial distribution of the dineutron can be investigated by the neutron knock-out reaction.
The missing momentum, which is the momentum of the knock-out neutron in the $^{11}$Li nucleus, can be determined experimentally assuming the impulse approximation.
Then the spatial distribution of the dineutron can be interpreted through the Fourier transformation.

\begin{figure}[htp]
  \includegraphics[width=1.0\columnwidth, angle=0, bb=2 5 240 180,clip]{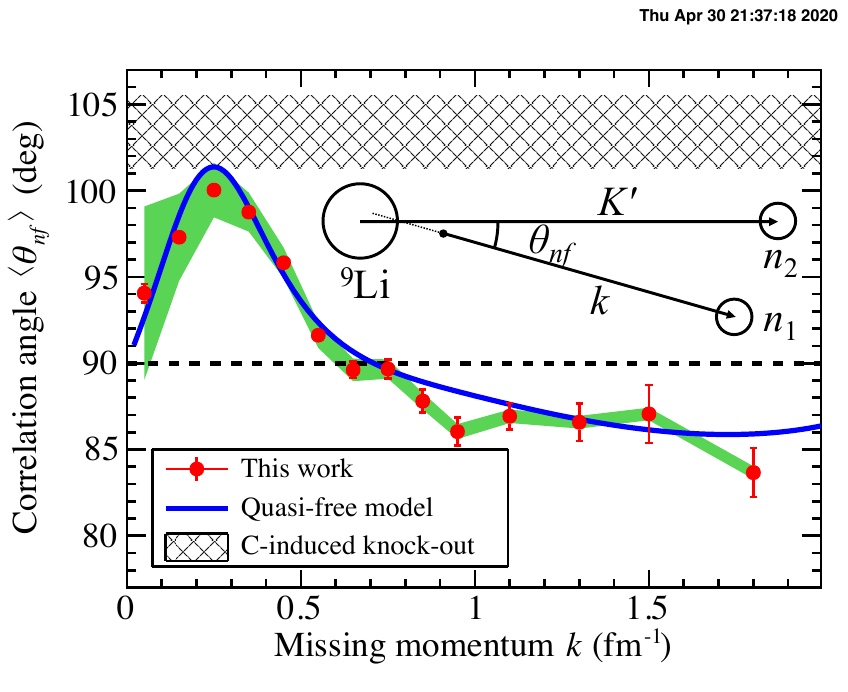}
  \caption{Mean values of the correlation angle $\langle\theta_{nf}\rangle$ in momentum space.
    Red points show the data in this study.
    Error bars and green line show the statistical and systematic uncertainty, respectively.
    Blue curve shows the quasi-free model calculation.
    Black hatched area shows the average correlation angle obtained in a previous study~\cite{PhysRevLett.83.496}.
    Black dashed line shows the expected $\langle\theta_{nf}\rangle$ value for the two uncorrelated neutrons.
    Inset shows a schematic diagram of the correlation angle $\theta_{nf}$ in $^{11}$Li.
    Black dot on the line between $^9$Li and $n_1$ represents the center of mass of $^{11}$Li.
    }
  \label{fig:thy}
\end{figure}

The neutron knock-out reaction was pioneered in experiments at GSI Helmholtz Centre for Heavy Ion Research (GSI) at a high energy (300~MeV/nucleon) with a carbon target~\cite{PhysRevLett.83.496,Simon2007267}.
The correlation angle between the valence neutrons $\theta_{nf}$ is defined for momentum space in the so-called Y-type Jacobi coordinates~\cite{Betounes2001}.
Schematic diagram of $\theta_{nf}$, which is comparable with $180^\circ - \theta_{12}$, is shown later in the inset of Fig.~\ref{fig:thy}.
Thus, the spatially compact dineutron should show an angle greater than 90$^\circ$.
Indeed, the GSI experiment exhibited an enhancement at $\theta_{nf}>90^\circ$, providing evidence for the existence of the dineutron in $^{11}$Li.
Another study constructed an analytical wave function to reproduce these data, and from the wave function, the averaged opening angle $\langle\theta_{12}\rangle$ was deduced to be 61.7$^\circ$ for the $J=0$ pair~\cite{Shulgina2009175}.
This result is consistent with the reevaluated results~\cite{PhysRevC.76.051602,PhysRevC.76.047302,Myo2008119561} of the $E1$ measurement~\cite{PhysRevLett.96.252502}.
Although the GSI experiment also showed that the slope of the $\cos\theta_{nf}$ distribution depends on the relative energy of $^{10}$Li ($^9$Li-$n$)~\cite{Simon2007267}, the dependence was not discussed in detail.

Here, we report the kinematically complete measurement of $^{11}$Li($p,pn$)$^{10}$Li in inverse kinematics with an unprecedentedly high statistics at 246~MeV/nucleon.
The nucleon target allows the whole volume of $^{11}$Li to be probed with the least effect of the absorption~\cite{Kikuchi01102016}.
In addition, the flexibility in kinematics can realize the quasi-free condition where the knock-out neutron is free from the FSI from $^{10}$Li.
Thus, the initial momentum of the neutron before the knock-out can be derived from the measured momenta of the recoil proton and knock-out neutron~\cite{RevModPhys.45.6}.
This measurement was made possible due to the unique combination of a highly-intense $^{11}$Li beam at the RI Beam Factory (RIBF) at RIKEN and a thick liquid hydrogen target MINOS device~\cite{Alexandre2011epj,Alexandre2013epj}.

The experiment was performed at RIBF, which is operated by RIKEN Nishina Center and the Center for Nuclear Study, The University of Tokyo.
A $^{11}$Li secondary beam was produced as a cocktail beam, which included $^{14}$Be and $^{17}$B, through a projectile fragmentation reaction from a $^{48}$Ca beam at 345~MeV/nucleon bombarding $^9$Be as the primary target.
The averaged beam intensity on the target was 400~pnA.
The secondary beam was selected and purified using the BigRIPS fragment separator~\cite{4277534}.
The average $^{11}$Li beam energy was 246~MeV/nucleon and the typical intensity was $1\times10^5$ particles per second with a purity of 70\%.
$^{11}$Li beam particles were identified on an event-by-event basis.

\begin{figure}[htp]%[htp!]
\begin{center}
  \includegraphics[width=1.0\columnwidth, clip, bb=40 75 620 410]{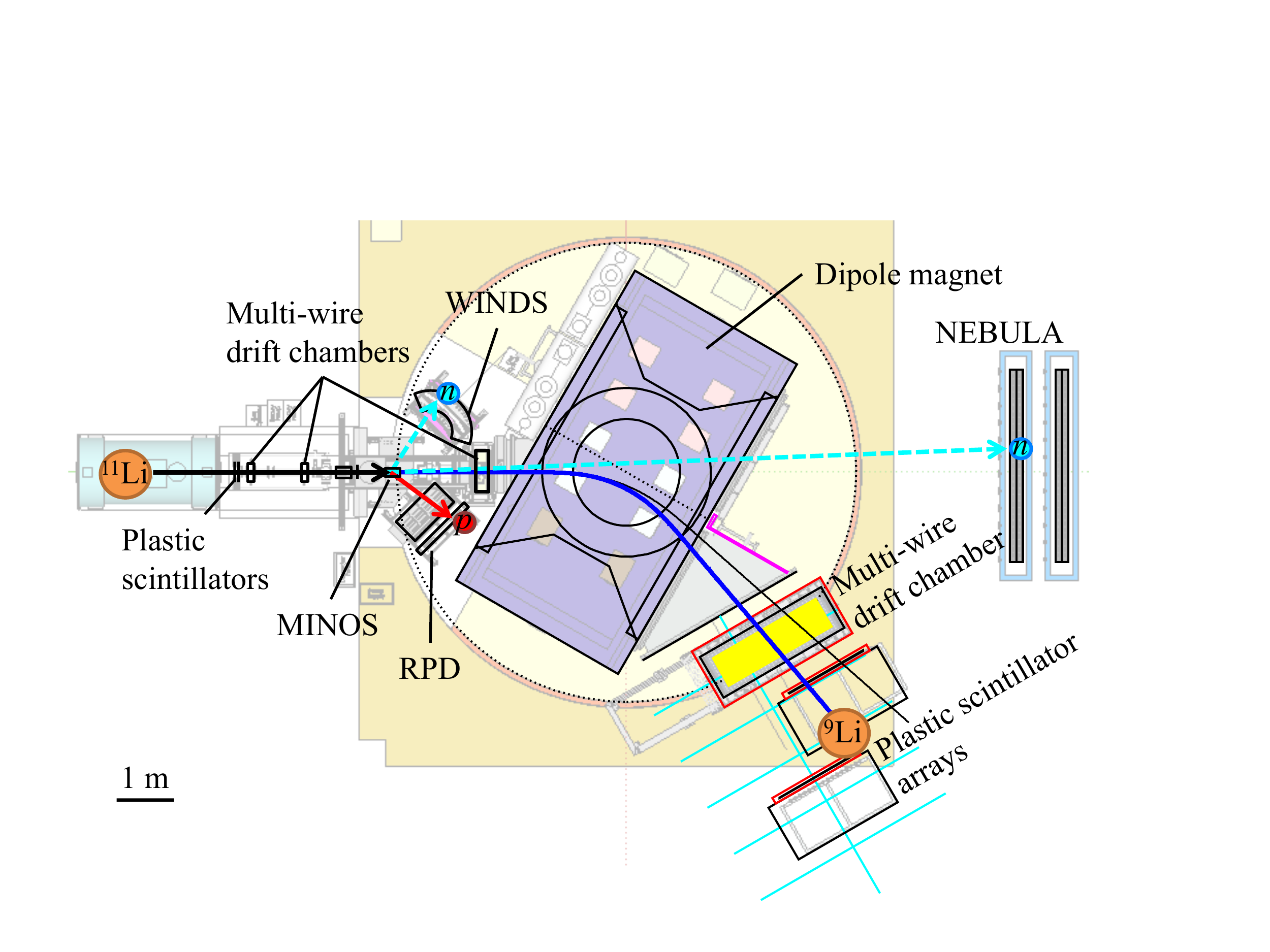}
\end{center}
\caption{Schematic of the setup. Arrows denote particle trajectories.}
\label{fig:setup}
\end{figure}

Figure~\ref{fig:setup} shows the main elements of the experimental setup around the secondary target.
The same setup was used in Ref.~\cite{CORSI2019134843}.
To achieve a high luminosity without degrading the resolution, the MINOS device, which was composed of a 15-cm-thick liquid hydrogen target coupled to a cylindrical time projection chamber (TPC) to determine the reaction vertex~\cite{SANTAMARIA2018138}, was installed.
The reaction vertex in the target was determined with an uncertainty of 6~mm at the full width at half maximum (FWHM) using a combination of the trajectory information of the incident $^{11}$Li beam and the recoil proton.

After a $(p,pn)$ reaction, the target proton (recoil proton) and one of the valence neutrons of $^{11}$Li (knock-out neutron) were scattered at large polar angles centered at approximately 45$^\circ$.
Then a recoil particle detector (RPD) was installed on the right-hand side of the beam line to detect the recoil proton scattered to $\theta_p = 30$--65$^\circ$ in the laboratory frame.
A large momentum transfer of $>1.5$~fm$^{-1}$, which is much larger than the typical two-neutron momentum in $^{11}$Li of 0.1~fm$^{-1}$~\cite{TANIHATA1992307}, was selected to ensure a clean quasi-free knock-out reaction condition.
The RPD was composed of a multi-wire drift chamber and a plastic scintillator hodoscope, which measured the scattering angle and the time of flight, respectively.
The momentum vector of the recoil proton at the reaction vertex was reconstructed by considering the energy loss.
The neutron detector array WINDS~\cite{Yasuda2016393} was installed on the left-hand side of the beam line ($-60^\circ<\theta_n<-25^\circ$).
WINDS measured the scattering angle and the time of flight of the knock-out neutron.
An unbound reaction residue $^{10}$Li was emitted in the very forward direction but immediately decayed into the heavy fragment $^9$Li and another neutron.
These momenta were analyzed by a SAMURAI spectrometer~\cite{Kobayashi2013294,Shimizu2013739} and a neutron detector array NEBULA~\cite{Nakamura2016156}, respectively.

The reaction channel of $^{11}$Li$(p,pn)^{10}$Li$^*$$\rightarrow$$^9$Li$+n$ was identified by detecting all particles in the final state.
The missing momentum $\bm{k}$, which is the initial momentum vector of the knock-out neutron in the beam rest frame, is derived as
\begin{align}
  \bm{k} &:= \bm{k}_{\bm{n1}} = \bm{k'}_{\bm{n1}} + \bm{k'}_{\bm{p}} - \bm{k}_{\bm{p}}, \label{eq:initial_momentum}
\end{align}
where $\bm{k_i}$ and $\bm{k'_i}$ represent the momentum vectors of particle $i$ [$n1$: knock-out neutron, $p$: target (recoil) proton] in the initial and final states, respectively.

Following Refs.~\cite{PhysRevLett.83.496,Simon2007267}, the correlation angle $\theta_{nf}$ is defined as
\begin{align}
  \cos\theta_{nf} &= \frac{\bm{K'} \cdot \bm{k}}{|\bm{K'}| |\bm{k}|}, \label{eq:thY_def}\\
  \bm{K'} &= \bm{k'}_{\bm{n2}} - \bm{k'_f}, \label{eq:KY_def}
\end{align}
where $\bm{k'}_{\bm{n2}}$ and $\bm{k'_f}$ represent the momentum vectors of the decay neutron and the heavy fragment $^9$Li, respectively.

The relative energy $E_\mathrm{rel}$ was obtained by subtracting the sum of the decay neutron and the heavy fragment $^9$Li masses from the invariant mass of $^{10}$Li.

The acceptance and efficiency of the setup were evaluated by \textsc{Geant4}~\cite{Agostinelli2003250,1610988} simulations.
The overall acceptance is 0.6\%, which is limited mainly by the azimuthal angle coverage of RPD and WINDS.
The average resolutions on the missing momentum $k$ and the correlation angle $\theta_{nf}$ are 0.17~fm$^{-1}$ (FWHM) and 12$^\circ$ (FWHM), respectively.

\begin{figure}[htp]
  \centering
  \includegraphics[width=1.0\columnwidth, angle=0, bb=10 0 244 228,clip]{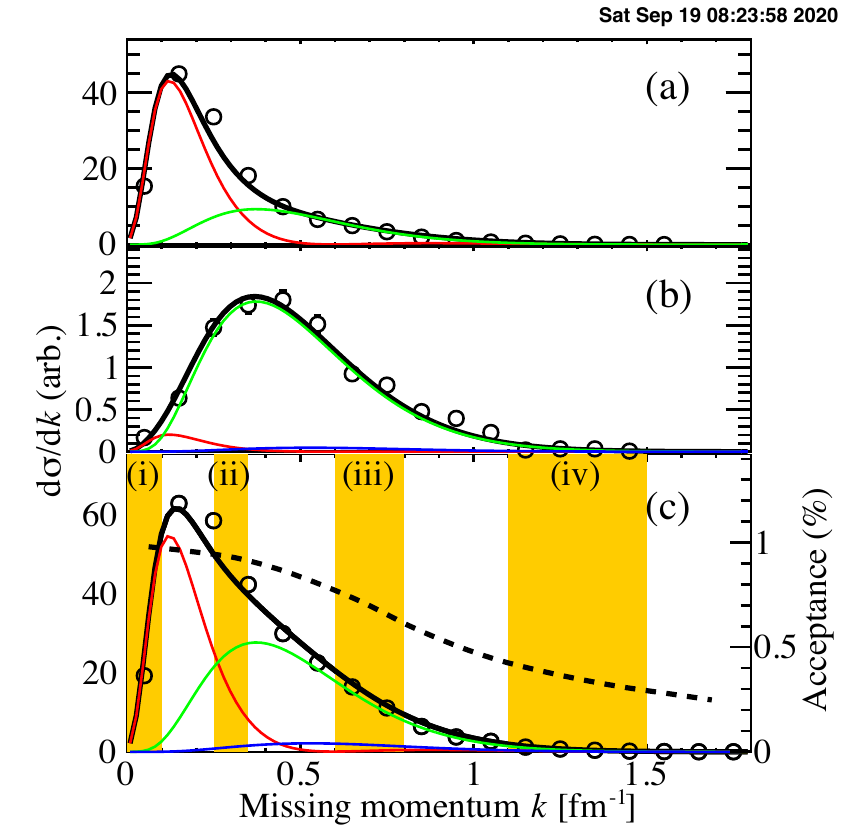}
  \caption{Missing momentum $k$ distribution for (a) $0\leq E_\mathrm{rel}<0.5$~MeV, (b) $2.0\leq E_\mathrm{rel}<3.0$~MeV, and (c) all data.
            Horizontal and vertical axes show the missing momentum of the knock-out neutron and the differential cross section, respectively.
            Black dotted curve represents the experimental acceptance.
            Red, green, and blue solid curves represent DWIA calculations for $s$-, $p$-, and $d$-waves, respectively.
            Black thick solid lines represent the fitting result.
            Four orange colored areas (i)--(iv) represent regions with characteristic angular momentum distributions. See text for details.}
          \label{fig:erelky}
\end{figure}

Figure~\ref{fig:erelky} shows the cross section as a function of $k$ for three different ranges of the relative energy $E_\mathrm{rel}$.
The theoretical curves show that the calculated distribution shapes for different angular momentum components are quite distinct.
Hence, the calculations according to the distorted-wave impulse approximation (DWIA)~\cite{Kikuchi01102016} can be fitted to the measured $k$ distribution to determine each multipole component.
Table~\ref{tb:multipole_fractions} summarizes the two-neutron configurations of $(1s)^2$, $(0p)^2$, and $(0d)^2$ considered for the fitting and their integrated fractions.
Including the contributions from $f$- and higher orbits did not change the fitting result.

The fitting was performed for each $^9$Li$+n$ relative energy bin.
Figures~\ref{fig:erelky}(a) and (b) show the results of the fitting for $0 \le E_{\rm rel} < 0.5$~MeV and  $2.0 \le E_{\rm rel} < 3.0$~MeV where the $1s$- and $0p$-components dominate, respectively.

The red, green, and blue curves in Fig.~\ref{fig:erelky}(c) are the fitted results for the $1s$, $0p$, and $0d$ components, respectively.
Assuming the $J=0$ pair of two valence neutrons, the integrated fractions of the two-neutron configurations $(1s)^2$, $(0p)^2$, and $(0d)^2$ are 35${}\pm{}$4\%, 59${}\pm{}$1\%, and 6${}\pm{}$4\%, respectively.
The error is attributed to the limited information on the $^9$Li-$n$ optical potential at energies above 100~MeV.
Our results confirm comparable contributions from $s^2$ and $p^2$, which are accompanied with a small $d^2$ contribution.
Table~\ref{tb:multipole_fractions} compares the values in this study to those from previous experimental and theoretical results.

\begin{table*}[htp]
  \centering
  \caption{Comparison of the integrated fraction for each multipole in percentage (\%) of experimental and theoretical studies.}
          \label{tb:multipole_fractions}
          % \begin{tabular}{llrrrrr}
          \begin{tabular}{lllcccccc}
            \hline
            \hline
            &          & & $(1s_{1/2})^2$ & $(0p_{3/2})^2$ & $(0p_{1/2})^2$ & $(0d_{5/2})^2$ & $(0d_{3/2})^2$ \\
            % &           & $(s_{1/2})^2$ & $(p_{3/2})^2$ & $(p_{1/2})^2$ & $(d_{5/2})^2$ & $(d_{3/2})^2$ \\
            \hline
            \\[-0.7cm]
            \multirow{9}{*}{Exp.}
            % & & & & & \multicolumn{2}{c}{\vspace{-0.4cm}$\underbrace{\qquad\qquad}_{\displaystyle 6\pm4}$} \\
            % & \vspace{-0.4cm} & & \multicolumn{2}{c}{$\underbrace{\qquad\qquad}_{\displaystyle 59\pm1}$} & \multicolumn{2}{c}{$\underbrace{\qquad\qquad}_{\displaystyle 6\pm4}$} \\[-2ex]
            & & & & \multicolumn{2}{c}{$\underbrace{\qquad\qquad}_{\displaystyle 59\pm1}$} & \multicolumn{2}{c}{$\underbrace{\qquad\qquad}_{\displaystyle 6\pm4}$} \\[-0.4cm]
            % \hline
            & This work; quasi-free $(p,pn)$ & 
            % & 35${}\pm{}$4  &    0$^a$ & 59${}\pm{}$1  &               &               \\
            & 35${}\pm{}$4  &    &  &               &               \\
            & C-induced knock-out & \cite{Simon2007267}
            & 45${}\pm{}$10 &     3--5 & 45${}\pm{}$10 &  \multicolumn{2}{c}{10${}\pm{}$8}               \\
            & Detailed analysis of Ref.~\cite{Simon2007267} & \cite{Shulgina2009175}
            % & 36            & 8        & 55            &               &               \\
            & 36.8          & 9.9        & 46.8           &               &               \\
            & $(p,pn)$ & \cite{Aksyutina20131309}
            &               &          &               &  \multicolumn{2}{c}{11${}\pm{}$2}               \\
            & $(p,d)$ & \cite{Sanetullaev2016481}
            & $\geq{}$44       &          & 33${}\pm{}$12 &            &               \\
            & $(p,t)$ & \cite{PhysRevLett.100.192502}
            & 31--45        &          & 51--64        &               &               \\
            % & Compilation~\cite{Shulgina2009175}
            % & \del{36}\add{36.8}          & \del{8}\add{9.9}        & \del{55}\add{46.8}           &               &               \\
            \hline
            \multirow{4}{*}{Theor.}
            & Few-body & \cite{PhysRevC.72.044321}
            &               &          &          59.1 &               &               \\
            & Coupled-channel & \cite{PhysRevC.87.034606}
            &          44.0 &      2.5 &          46.9 &           3.1 &            1.7 \\
%            & Ref.~\cite{PhysRevC.91.017303}
%            & $33^{+3}_{-5}$& \multicolumn{2}{c}{\vspace{-0.4cm}$\underbrace{\qquad\qquad}_{\displaystyle 34\pm6}$} & \multicolumn{2}{c}{$\underbrace{\qquad\qquad}_{\displaystyle 15\pm2}$} \\
%            & Ref.~\cite{FORTUNE2016577}
%            &      $51\pm5$ &          &               &               &                \\
            & Tensor-optimized shell model & \cite{Ikeda2010}$^*$
            &          46.9 &      2.5 &          42.7 &           4.1 &            1.9 \\
            & Transfer to the continuum & \cite{GOMEZRAMOS2017115}
            &            67 &          &            31 &             1 &                \\
            \hline
            \hline
          \end{tabular}
              {\flushleft \small
                % $^a$ Fixed. \\
                %$^b$ 0.6\% and 0.5\% contributions for $(f_{7/2})^2$ and $(f_{5/2})^2$ in Ref.~\cite{Ikeda2010}. \\
                $^*$ 0.6\% and 0.5\% for $(f_{7/2})^2$ and $(f_{5/2})^2$, respectively. \\
              }
\end{table*}

To simplify the argument, here, we concentrate on the ratio of the $s^2$ and $p^2$ components.
The $s/p$ ratio is 0.59, which is consistent with the analysis using the multi-step transfer calculation of the $^{11}$Li$(p,t)$ measurement at 3~MeV/nucleon~\cite{PhysRevLett.100.192502}.
a larger $s/p$ value of $\geq 1.3$ has been reported for $^{11}$Li$(p,d)$ measured at 6~MeV/nucleon~\cite{Sanetullaev2016481}.
However, the $s$-wave fraction was not directly determined from the experimental data, and the estimate may have a large uncertainty.
The knock-out measurement with a carbon target at 260~MeV/nucleon shows $s/p \sim 1$~\cite{Simon2007267} but with the error is on the order 40\%.
This is marginally consistent with the present result of 0.59 within a 1$\sigma$ variation.
However, it should be noted that the strong peripherality of the heavy-ion-induced knock-out reaction employed in Ref.~\cite{Simon2007267} may underestimate the $p$-wave, which has a larger fraction in the inner part of $^{11}$Li than the $s$-wave.
This may overestimate the $s/p$ ratio.
The effects of the different peripheralities in the $(p,pn)$ and the heavy-ion-induced knock-out reactions will be discussed later.

The experimental $s/p$ ratio was compared to the theoretical calculations.
A three-body model calculation with a density dependent force predicts an $s/p$ ratio of less than 0.7~\cite{PhysRevC.72.044321}.
Although the $s$-wave fraction was not explicitly shown, the result is consistent with the present result.
The tensor-optimized shell model, which considers the tensor and pairing correlations in the $^9$Li core, predicts an $s/p$ ratio of 0.9--1.1~\cite{PhysRevC.87.034606,Ikeda2010}.
Applying the transfer-to-the-continuum reaction framework to the $^{11}$Li$(p,pn)$ data~\cite{Aksyutina2008430} gives an $s/p$ ratio as large as 2.2~\cite{GOMEZRAMOS2017115}.
The reason that the $s/p$ ratio differs among theoretical models is a topic for future studies.

\begin{figure}[htp]
  \centering
  \includegraphics[width=1.0\columnwidth, angle=0, bb=2 5 240 185, clip]{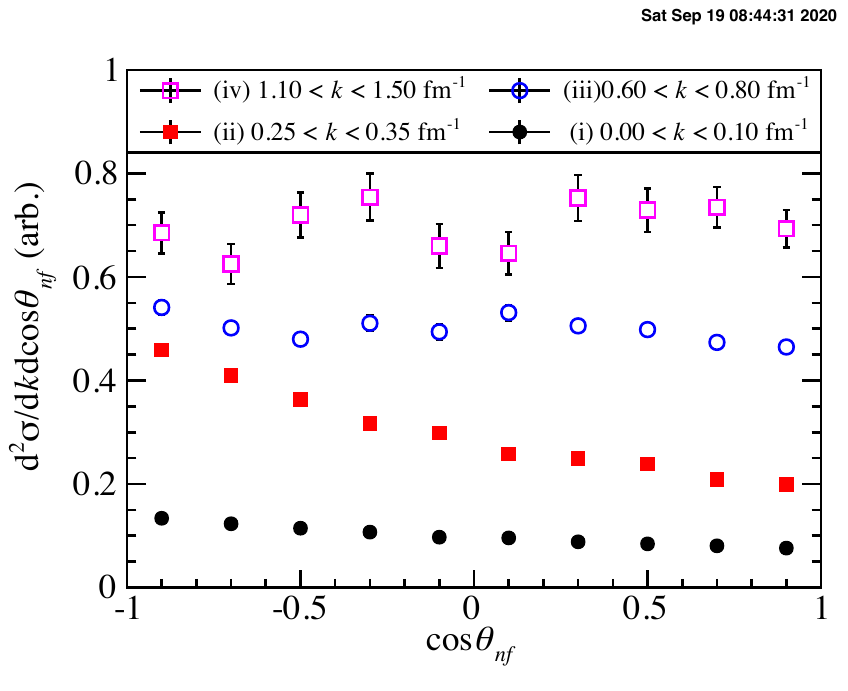}
  \caption{$\cos\theta_{nf}$ distribution for the different intervals (i)--(iv) of the missing momentum $k$.
    Horizontal and vertical axes show the correlation angle and double differential cross section, respectively.
    Vertical axis of each spectrum is scaled for comparison.
    Error bars represent the statistical uncertainty.
  }
  \label{fig:angplot}
\end{figure}

Figure~\ref{fig:angplot} shows the correlation-angle distributions at different $k$.
There is an apparent $k$-dependence.
Herein we picked up four $k$ regions (i)--(iv).
% The $s$- and $p$-wave contributions were obtained by the aforementioned multipole decomposition, as shown in Fig.~\ref{fig:erelky}(c).
% At (ii) $k={}$0.25--0.35~fm$^{-1}$, where the $s$- and $p$-wave fractions are similar, the distribution shows a strong enhancement in $\cos\theta_{nf}<0$ ($\theta_{nf}>90^\circ$), providing evidence of a dineutron correlation.
At (ii) $k={}$0.25--0.35~fm$^{-1}$, where the $s$- and $p$-wave fractions obtained by the aforementioned multipole decomposition are similar, the distribution shows a strong enhancement in $\cos\theta_{nf}<0$ ($\theta_{nf}>90^\circ$), providing evidence of a dineutron correlation.
The trend is less prominent at the lower (i: 0.0--0.1~fm$^{-1}$) and higher (iii: 0.6--0.8~fm$^{-1}$) $k$ regions, where one angular momentum is dominant.
% In contrast, $\cos\theta_{nf}$ is almost consistent with 0 at (iv) $k={}$1.1--1.5~fm$^{-1}$ where the $d$ wave appears.
In contrast, the distribution has no apparent dependence of $\cos\theta_{nf}$ at (iv) $k={}$1.1--1.5~fm$^{-1}$ where the $d$ wave appears.
Below, the mean value of the correlation angle $\langle\theta_{nf}\rangle(k)=\int\theta_{nf} P(\cos\theta_{nf},k)\, \d\cos\theta_{nf}$, where $P(\cos\theta_{nf},k)$ is a normalized $\cos\theta_{nf}$ distribution, was used to discuss the $k$-dependence of the dineutron correlation.

Figure~\ref{fig:thy} illustrates a clear $k$ dependence of $\langle\theta_{nf}\rangle$.
$\langle\theta_{nf}\rangle$ takes a maximum value of $\sim100^{\circ}$ at $k\sim 0.3$~fm$^{-1}$.
The correlation angle of $103.4\pm2.1^\circ$ obtained with the knock-out reaction by a carbon target~\cite{PhysRevLett.83.496} was averaged over the whole momentum range.
However, it is even larger than the maximum $\langle\theta_{nf}\rangle$ value in the present work.
This discrepancy may be due to the different peripheralities of the probes.
The carbon target used in Ref.~\cite{PhysRevLett.83.496} selectively probed the surface of $^{11}$Li, where the dineutron correlation is favored.
Theoretical models of the knock-out process, including the probe transparency, should realize a quantitative comparison of these results.

The mean correlation angle $\langle\theta_{nf}\rangle$ decreases at smaller and larger $k$ values.
It crosses 90$^\circ$ at $k\sim0.6$~fm$^{-1}$, suggesting that the dineutron is not present in the tail of the halo or in the inner part of $^{11}$Li.

The increase of the $\langle\theta_{nf}\rangle$ value in the limited region may be a signature of radial localization of the dineutron in $^{11}$Li around the $^9$Li core.
With the help of the quasi-free model described below, the peak structure of $\langle\theta_{nf}\rangle$ at $k\sim0.3$~fm$^{-1}$ can be interpreted as the dineutron correlation, which is maximized at $r\sim3.6$~fm from the center of the $^9$Li core.
This is consistent with the three-body model calculation that includes many-body correlations~\cite{PhysRevLett.99.022506} where the root-mean-square radius between two neutron in $^{11}$Li takes a minimum at $r\sim3.2$~fm.

The quasi-free model, which consists of a combination of the $^9\mathrm{Li}+n+n$ three-body model and the knock-out reaction model in Ref.~\cite{Kikuchi01102016}, well reproduces the obtained $k$-dependence of the correlation angle.
The three-body model considered the contributions from the excited $^9$Li core via the coupled-channel calculation~\cite{PhysRevC.87.034606}.
The folding potential of the effective $NN$ potential with the $^9$Li core density was used for the $^9$Li-$n$ interaction and its parameter was modified to reproduce the observed two-neutron separation energy.
For the $n$-$n$ interaction, the realistic Argonne $v8'$ force~\cite{PhysRevC.51.38} was used.
To describe the knock-out process, the final scattering states were approximated with the products of the plane wave of the knock-out neutron and the remaining $^9\mathrm{Li}+n$ system.
The FSI in the $^{10}$Li resonance was also taken into account~\cite{Kikuchi01102016}.
The absorption effect by the proton target was considered in a manner similar to that in Ref.~\cite{Kikuchi01102016} except that the same function form of the damping factor and its parameters were modified to reproduce the observed value of $\langle\theta_{nf}\rangle$ in this experiment since the optical potential between $^9$Li and the proton target is unknown.
The successful reproduction of the experimental data with the reaction model highlights the simple mechanism of the quasi-free $(p,pn)$ reaction, which enables the radial-dependent properties of dineutron in nuclei to be extracted.

It is interesting to compare the results with the theoretical predictions on the dineutron correlation in infinite nuclear matter.
The present analysis implies that the dineutron correlation is prominent only around the $^9$Li core surface where the density is $10^{-3}\lesssim\rho/\rho_0\lesssim10^{-2}$, and it becomes weaker at the tail of the halo where the density is extremely low.
Hartree-Fock-Bogoliubov calculations~\cite{PhysRevC.73.044309} predict a similar density dependence as the dineutron develops in a density region of $10^{-4}\lesssim\rho/\rho_0\lesssim0.5$ and vanishes at lower and higher densities.
If this is a universal characteristic of the dineutron correlation, it should exist at the low-density surface of any neutron-rich nuclei.
Actually, the three-body model calculation~\cite{PhysRevLett.99.022506} predicts that the dineutron is formed in a limited region around the surface of $^6$He, $^{16}$C, and $^{24}$O.
Future $(p,pn)$ experiments should reveal the nature of the dineutron correlation in these nuclei.

In summary, herein the dineutron correlation in $^{11}$Li was investigated via the quasi-free $(p,pn)$ knock-out reaction.
High-statistics data were acquired using the high-intensity $^{11}$Li beam at the RIBF and the MINOS device, enabling a detailed analysis of the $^{11}$Li structure.
The fraction of each two-neutron configurations is determined as 35${}\pm{}$4\% $(1s)^2$, 59${}\pm{}$1\% $(0p)^2$, and 6${}\pm{}$4\% $(0d)^2$.
The correlation angle $\cos\theta_{nf}$ distribution has an asymmetric shape and a missing momentum $k$ dependence, indicating that the dineutron correlation is localized radially on the $^{11}$Li surface.
The $k$-dependence of the $\langle\theta_{nf}\rangle$ is well reproduced by the quasi-free model calculations.
The dineutron in $^{11}$Li is localized at $k\sim0.3$~fm$^{-1}$, corresponding to the nuclear surface $r\sim3.6$~fm.
This behavior is consistent with the Hartree-Fock-Bogoliubov calculation for infinite nuclear matter.

\begin{acknowledgments}
  We extend our appreciation to the RIKEN Nishina Center, the Center for Nuclear Study, The University of Tokyo, and the accelerator staff for their efforts in delivering an intense beam.
  We also would like to thank K.~Hagino for the helpful discussion.
  The development of MINOS and the core MINOS team were supported by the European Research Council through the ERC (Grant No. MINOS-258567).
  This work was supported in part by KAKENHI (Grant No. JP16K05352, JP16H02179, and JP18H05404).
  D.~K. acknowledges support from the Rare Isotope Science Project of Institute for Basic Science funded by the Ministry of Science and ICT and NRF of Korea (numbers NRF-2013M7A1A1075764 and NRF-2019R1I1A1A01058354).
  L.~S. acknowledges support from the Institute for Basic Science (IBS-R031-D1).
  J.~G., F.~M.~M., and N.~A.~O. acknowledge partial support from the French-Japanese LIA-International Associated Laboratory for Nuclear Structure Problems as well as the French ANR14-CE33-0022-02 EXPAND.
\end{acknowledgments}

%\bibliography{dissertation_mod}

\end{document}